\documentstyle[aps,prl,epsf,multicol]{revtex}
\newcommand{\be}{\begin{equation}}
\newcommand{\ee}{\end{equation}}
\newcommand{\ba}{\begin{array}}
\newcommand{\ea}{\end{array}}
\newcommand{\bea}{\begin{eqnarray}}
\newcommand{\eea}{\end{eqnarray}}

\newcommand{\ra}{\rightarrow}

\begin{document}
\draft
\title{Dynamical surface structures in multi-particle-correlated
surface growths }
\author{Yup Kim$^{1,2}$, T. S. Kim$^1$ and Hyunggyu Park$^3$}
        
\address{$^1$ Department of Physics and Research Institute of Basic Sciences, Kyung Hee University,
         Seoul, 130-701, Korea}
\address{$^2$ Asia Pacific Center for Theoretical Physics, Korea}
\address{$^3$ Department of Physics, Inha University,
         Inchon 402-751, Korea}
\date{\today}
\maketitle

\begin{abstract}
We investigate the scaling properties of the interface fluctuation width 
for the $Q$-mer and $Q$-particle-correlated deposition-evaporation models. 
These models are constrained with a global conservation law that the particle
number at each height is conserved modulo $Q$. In equilibrium, the stationary 
roughness is anomalous but universal with roughness exponent $\alpha=1/3$,
while the early time evolution shows nonuniversal behavior with growth
exponent $\beta$ varying with models and $Q$. Nonequilibrium surfaces display
diverse growing/stationary behavior. The $Q$-mer model shows a faceted structure,
while the $Q$-particle-correlated model a macroscopically grooved structure.
\end{abstract}


\begin{multicols}{2}

\section{Introduction}

Since the dynamical scaling theory for  kinetic surface 
roughening \cite{FAV} was suggested, numerous surface growth models 
have been studied \cite{dyn}. 
The interface fluctuation width $W$ developed during time $t$, starting from
the flat surface, follows the scaling ansatz
\be 
W(L,t)=L^\alpha f({t \over L^{z_W}}), \label{scaling} 
\ee 
where $L$ is linear size of the substrate. The scaling function 
$f(x)\ra {\rm const.}$ for $x\gg 1$ and 
$f(x) \sim x^\beta$ $(\beta =\alpha/z_W)$ for $x\ll 1$\cite{FAV,dyn}. 
One-dimensional (1D) models with conventional local growth dynamics display the
universal roughness exponent $\alpha=1/2$, as exemplified in
the Kadar-Parisi-Zhang (KPZ) universality class \cite{KPZ} and 
the Edwards-Wilkinson (EW) class \cite{EW}. 

A 1D surface can be mapped on the time trajectory of a particle in 1D by
identifying the height $h(x)$ at each column $x$ with the particle position
$n(t)$ at time $t=x$. The step heights are bounded and
uncorrelated beyond a finite distance in the substrate direction.
This implies that the particle performs a random walk without 
long-range diffusion and long-range temporal correlations, which
is the characteristic of the normal random walks.
Therefore the interface width scales  with exponent $\alpha=1/2$,
corresponding to the dispersion of normal random walks: $\Delta n(t) \sim t^{1/2}$.

Recently, Noh, Park, and den Nijs \cite{NPD} introduced a dissociative dimer 
deposition-evaporation model with restricted solid-on-solid (RSOS) constraint.
The surface grows and erodes by the deposition and evaporation of dimers 
aligned with the surface. These dimers dissociate on the surface such that 
evaporating dimers do not necessarily consist of original partners.
They found that the equilibrium surface is anomalously rough with
$\alpha = 0.29(4)$ and conjectured the exact value of $\alpha=1/3$ by
exploring a close relationship between the dimer model and the even-visiting 
random walk model \cite{NPD,NKPD}. 
They also argued that the anomalous value of $\alpha=1/3$ 
is universal in generalized $Q$-mer type equilibrium models $(Q\ge2)$. 

The $Q$-mer aspect requires that the number of particles at each surface height level must 
be conserved modulo $Q$. The dissociative nature translates it into a nonlocal global constraint. 
So 1D surfaces of $Q$-mer models correspond to time trajectories of a random walker 
with a global constraint that the walker should visit each site (height level) $Q$ multiple times
before it terminates. The dispersion of $Q$-visiting random walkers (QVRW) is much more suppressed
than the normal ones and behaves asymptotically as $\Delta n(t) \sim t^{1/3}$ for $Q\ge 2$ 
\cite{NKPD,CCM,Luck}.
This leads to the conjecture that the roughness exponent is $\alpha=1/3$ for general $Q$-mer 
type equilibrium models except for the monomer limit ($\alpha=1/2$ for $Q=1$).

However, the QVRW ensemble is not equivalent to the ensemble generated by the $Q$-mer dynamics.
In fact, there exist infinitely many constants of motions in the $Q$-mer dynamics.
For example, the $Q$-mer dynamics conserves a local-type quantity. 
Upon deposition/evaporation of a $Q$-mer, 
the surface heights at $Q$ consecutive columns change by one unit simultaneously.
It leads to conservation of $k=2\pi /Q$ component of the Fourier transformed surface height, defined as
${\tilde h}(k) \equiv \sum_{x=1}^L  \exp {[-ikx]} h(x)$.
Surface configurations with different values of ${\tilde h}(k=2\pi/Q)$ are dynamically disconnected
by the $Q$-mer dynamics.  
The phase space for the QVRW ensemble is divided into dynamically disconnected sectors,
labeled by ${\tilde h}(k=2\pi/Q)$. However, there is much more complex structure
inside each sector. Each sector decomposes into infinitely many disconnected subsectors,
which can be characterized by a nonlocal construct, i.e.~so called irreducible string \cite{BD}. 
Therefore, the $Q$-mer dynamics is strongly non-ergodic and 
the scaling property may be sector-dependent. 

Starting from a flat surface of size $L$ as multiples of $Q$,
one can explore only a subsector (null string) with ${\tilde h}(k=2\pi/Q)=0$.
Numerical results for dimer dynamics \cite{NPD} show strong finite size corrections
and the extrapolation to infinite size limit overshoots significantly 
the expected value of $\alpha=1/3$ \cite{NPD}, even though statistical errors are too large
to conclude whether this overshooting is real. 

This may be due to strong non-ergodic behavior of the $Q$-mer dynamics. 
In Ref. \cite{NKPD}, Noh {\em et al} allow diffusion of a single particle on a terrace, 
which restores broken ergodicity partially. Jumps across steps are forbidden.
A certain combination of deposition/evaporation and diffusion may be equivalent to 
a process of deposition/evaporation of $Q$ particles not necessarily at $Q$ consecutive columns
but randomly placed on a terrace. It clearly does not conserve ${\tilde h}(k=2\pi/Q)$
and opens some dynamic connection links between sectors.
Numerical results for $Q=2$ still show strong finite size corrections but the 
overshooting is weakened and the estimated asymptotic value of
$\alpha=0.31(3)$ is a little bit closer to the conjectured value of $1/3$ \cite{NKPD}.

In this paper, we introduce a new model that fully restores the ergodicity. 
We allow deposition/evaporations of $Q$ particles at randomly chosen $Q$ columns 
with equal surface heights. These columns do not necessarily share the same terrace. 
These processes still conserve the global mod $Q$ conservation
at each height level and open all possible dynamic connections between sectors. 
We call this model as $Q$-particle-correlated deposition-evaporation model. 
The ensemble generated by this new type dynamics should be
equivalent to the QVRW ensemble. 
As we shall see later, we find that the finite size corrections for $Q=2$ are still considerable 
but the overshooting is fairly reduced. Our estimate for $\alpha$ is now $\alpha\simeq 0.33(3)$, 
which is fully consistent with the QVRW conjecture. 

We also investigated the scaling properties for $Q$-mer models and $Q$-particle correlated models
for $Q=3$ and 4. Interestingly, the asymptotic scaling seems to set in much earlier than $Q=2$ models
and the estimated values of $\alpha$ are very close to $1/3$ in both models. 
These results may suggest that the phase space structures for $Q=2$ and $Q\ge 3$ bear 
an intrinsic difference like in the monolayer version of the $Q$-mer models~\cite{BD}. 

Time-dependent behavior in early time regime is much more complex and sector-dependent.
Various interesting dynamic properties have been reported by Grynberg in the body-centered 
solid-on-solid (BCSOS) version of $Q$-mer models~\cite{Gryn}. In this paper, we focus only on 
a subsector (null string) containing a flat surface. Our numerical results show that the scaling exponent 
$\beta$ in Eq.(\ref{scaling}) varies slightly but definitely with $Q$ in both $Q$-mer and 
$Q$-particle correlated models. Our estimated values 
of $\beta$ for the $Q$-mer models are consistent with those for the BCSOS version, but
different from those for the $Q$-particle correlated models. It may reflect non-ergodic nature
of the $Q$-mer dynamics. In contrast to the universal scaling property
in the steady state (universal $\alpha$), the dynamic properties characterized by $\beta$ 
vary with models and $Q$.

Excursion off equilibrium immediately changes the surface structure drastically \cite{NPD}.
Without balance between deposition and evaporation, the surface grows or erodes
indefinitely with time. 
In the $Q$-mer models, these nonequilibrium surfaces turn out to be always faceted 
due to spontaneous formation of pinning valleys or pinning hill tops \cite{NPD}. 
In finite systems, the surface displays a repeated jerky motion in time, i.e., alternating series of
locking in a facet shape and unlocking/evolving into another facet shape. 
Time scale of this jerky motion is dominated by duration of a facet shape, which 
diverges exponentially with system size. The global facet structure in the steady state
assures that the interface width $W$ scales as system size $L$ ($\alpha=1$). 
Along with the exponential time scale, one can predict that $W\sim \log t$ for early time regime.
These results are confirmed numerically.

In case of the $Q$-particle-correlated models, nonequilibrium surfaces are not sharply faceted but 
display a global grooved structure with considerable local fluctuations. This surface is similar
to those found in the conserved RSOS (CRSOS) model \cite{KIKJ}. The roughness exponent
is found to be $\alpha=1.00(5)$ for all $Q\ge 2$. Early time growth behavior is
governed by a power law with nonuniversal exponent $\beta$ varying with $Q$ and the
magnitude of deposition/evaporation bias.

Outline of this paper is as follows. 
In Sec.~II, we introduce the $Q$-particle-correlated models
as well as the $Q$-mer models. 
In Sec.~III and IV, we present numerical results for 
the scaling properties of the equilibrium and nonequilibrium surfaces, respectively. 
Similarities and differences between two models are discussed.
We conclude with a brief summary and discussion in Sec.~V.

\section{Models}

Consider the 1D surface configurations described by integer height
variables $h(x)=0,\pm1,\pm2, ...$ with $x=1, ... ,L$.
They are subject to the RSOS constraint, $h(x)-h(x+1)=0, \pm 1$, and
periodic boundary conditions, $h(x+L)=h(x)$. 

The growth rule for the $Q$-particle-correlated deposition-evaporation model
is as follows. First, select $Q$ columns $\{x_1, x_2, ...,x_Q \} $
randomly. These columns do not have to be adjacent to each other.
If the heights $\{h(x_j)\}$ of the selected columns are all equal, i.e., $h(x_1)= ...=h(x_Q)$, 
then simultaneous deposition of $Q$ particles, $h(x_i) \rightarrow h(x_i)+1$ for $i=1,..,Q$, 
is attempted with probability $p$, or simultaneous evaporation, $h(x_i) \rightarrow h(x_i)-1$ 
for $i=1,...,Q$, with probability $q=1-p$.  Any attempt is rejected if it would result in
violating the RSOS constraint. 
If any selected column height differs from other selected one, 
no dynamics occurs and  we select a new set of $Q$ columns randomly.
After $L$ such selection processes, Monte Carlo time $t$ is incremented by one unit. 

The equal height condition, $h(x_1)= ...=h(x_Q)$, constrains a set of selected columns
and generates correlated growth. Without this condition, any set of randomly selected
columns can evolve and the ordinary monomer-type RSOS model is recovered.
One can consider a more restrictive constraint on selected columns.
An interesting and limiting case is to require the immediate adjacency of selected columns.
This is the dissociative $Q$-mer deposition-evaporation model. 
In this $Q$-mer model, we attempt to select $Q$ consecutive columns only. 
The other evolution rules are equivalent to the $Q$-particle-correlated model. 
The generalization of both models to higher dimensions should be straightforward
and will be discussed elsewhere \cite{Next}. In this paper, we focus on 1D cases only.

\section{Equilibrium surfaces} 

With balance between deposition and evaporation processes ($p=q$), 
the detailed balance condition is satisfied. In the stationary state,
the Gibbs-type equilibrium surface is obtained. In this section, we investigate
the scaling property of the equilibrium surfaces for the $Q$-mer 
and $Q$-particle-correlated models and also their dynamic 
property approaching to equilibrium. We measure the interface 
fluctuation width $W$ as
\be 
W^2 (L,t)={1\over L} \sum_{x=1}^L \left\langle \left[ h(x,t)- 
{1\over L}\sum_{x=1}^L h(x,t) \right]^2 \right\rangle ,
\label{W2} 
\ee 
where $L$ is the substrate size.
We start with a flat surface of size $L$ as a multiple of $Q$, otherwise
the surface is forcefully pinned to the initial position.

\subsection{stationary state roughness}

For finite size scaling analysis, we run Monte Carlo simulations 
for system size  $L= 2^n$ $(n=4,...,10)$ in case of  $Q=2$ and 4,
and $L=2\times 3^n$ $(n=2,...,6)$ in case of $Q=3$. 
The interface width is measured and averaged over at least 100
independent Monte Carlo runs for each system size.
It grows in early time regime and saturates
to a finite value, which depends on size $L$. 
We again average over data in the saturated regime $(t\gg L^{z_W}$)
to estimate the stationary value of the width, $W(L,t=\infty)$. 

As explained in Eq. (\ref{scaling}), $W(L,t=\infty)$ scales as
$L^\alpha$ in the large $L$ limit. For efficient data analysis, 
we introduce effective exponents  $\alpha_{eff}$ as function of $L$
\be
\alpha_{eff}(L) = \frac{\ln W(2L, t=\infty)-\ln W(L, t=\infty)}{\ln (2L) - \ln L}.
\label{alpe}
\ee
The value of $\alpha$ can be retrieved by taking the $L\rightarrow\infty$ limit
for $\alpha_{eff}(L)$. 

Effective exponents for the $Q$-mer model are plotted against $1/L$ in Fig.~1.
For the dimer case, 
one can see usual $1/L$ type finite size corrections for small $L$, but
strong crossover corrections near $L=2^8$. 
The extrapolation of $\alpha_{eff}$ to the infinite size limit seems to overshoot 
significantly the expected value of $\alpha=1/3$.  However, the statistical
errors are too large to conclude that our numerical results exclude the possibility
of $\alpha=1/3$.  In fact, it is quite difficult to reduce statistical errors considerably
with moderate computing usage, because $W$ is very small even for large $L$
(less than 4 for $L=2^{10}$). We estimate $\alpha=0.29(4)$ for the dimer model. 
For $Q=3$ and 4, we do not find strong crossover corrections and 
our numerical data converge to $\alpha=1/3$ very nicely. 
We estimate $\alpha=0.32(4)$  for the trimer model and $\alpha=0.33(3)$ for 
the 4-mer model. 

Data for the $Q$-particle-correlated models are shown in Fig.~2. 
For $Q=2$, we find again strong crossover corrections, but the overshooting
is fairly reduced. We estimate $\alpha=0.32(3), 0.33(2)$ and 0.32(4) for 
$Q=2, 3$, and 4, respectively. These results are fully consistent with
the QVRW conjecture.

\subsection{Time dependent behavior}

In this subsection, we investigate the early time behavior
of the interface width. 
We measure the interface width for system size $L=10^4$ for $Q=2$ and 4,
and $L=2 \times 3^6$ for $Q=3$ up to $t=10^5 \sim 5\times 10^6$ 
and average over at least 100 independent samples.
In early time regime ($t\ll L^{z_W}$), it grows algebraically as
$W\sim t^\beta$, see Eq.(\ref{scaling}). To estimate the value of $\beta$,
we introduce effective exponents $\beta_{eff}$ as a function of $t$
\be
\beta_{eff}(t) = \frac{\ln W(t)-\ln W(t/10)}{\ln t - \ln (t/10)}.
\label{beta}
\ee
The value of $\beta$ can be retrieved from the long time limit for $\beta_{eff}$
within $t\ll L^{z_W}$, where finite size effects are negligible.

Effective exponents for the $Q$-mer model and the $Q$-particle-correlated model 
are plotted against $t$ in Fig.~3 and 4, respectively.
After initial transients, $\beta_{eff}$ become stabilized around an asymptotic value.
We estimate $\beta$ by averaging $\beta_{eff}$ in the stabilized regime. 
As seen in Fig.~3, the estimated values for $\beta$ are 0.108(1) for the dimer,
0.100(1) for the trimer, and 0.098(3) for the 4-mer model. 
Surprisingly, the value of $\beta$ decreases slightly but definitely with $Q$. 
Our finding is consistent with the result for the BCSOS type $Q$-mer model \cite{Gryn}.

The $Q$-particle correlated models show  even strong variance of $\beta$ on $Q$.
>From Fig.~4, we estimate $\beta=0.194(1)$, 0.140(2), 0.097(2) for $Q=2$, 3, and 4.
Our results suggest that $\beta$ continuously varies and 
approaches zero in the large $Q$ limit. 

The steady state scaling is quite universal and does not depend on 
$Q$$(\ge 2)$. This universality has been well established through a mapping to 
the $Q$-visiting random walks \cite{NKPD}. In contrast, our numerical results
strongly support nonuniversal dynamic scaling depending on $Q$.  
As yet, we do not have any reasonable explanation for this type of nonstandard scaling behavior.

It is not quite surprising that the $Q$-mer and $Q$-particle-correlated models display 
different dynamic scaling behaviors. The $Q$-mer dynamics are not ergodic, in contrast 
to the $Q$-particle-correlated dynamics. The non-ergodicity slows down the dynamics and 
the growth exponent should be smaller, which is consistent with our numerical finding. 

It is interesting to see that initial transients for the $Q$-mer and the $Q$-particle-correlated models
are completely upside down. We can give a reasonable explanation for this. 
In case of the $Q$-mer equilibrium model, it is clear that 
the early time behavior resembles the Edward-Wilkinson type (monomer model),
because the system does not feel the $Q$-mer aspect initially before pinning valleys/hilltops start to form. 
So, $\beta_{eff}$ should decrease from $\beta_{EW}=0.25$ to converge to the asymptotic value.
The $Q$-mer aspect should appear earlier for larger $Q$ with the same substrate size.
We expect that the initial transients survive for shorter period for larger $Q$, which can
be seen in Fig.~3.

In case of the $Q$-particle-correlated model, it is very difficult to grow the surface at the beginning.
As $Q$ particles deposit on randomly placed $Q$ sites, it is difficult to form aggregated islands
on which one can deposit more particles to grow the surface. So in the very early time regime, 
the surface would grow in a layer-by-layer fashion. So $\beta_{eff}$ will start from zero and
increase to the asymptotic value. Here, the initial transients last longer for larger $Q$. 
The rejection rate for the $Q$ particle deposition/evaporation
on the randomly selected sites increases exponentially with $Q$.
In fact, the initial transient behavior is so huge for $Q=4$ that  we use a rather small substrate size 
$L=10^3$ in Fig.~4.

\section{Nonequilibrium surfaces}

Without balance between deposition and evaporation ($p\neq q$), the surface may grow 
($p >1/2$) or erode ($p<1/2$) indefinitely with time, if one starts with a flat surface of size 
$L$ as a multiple of $Q$. Due to the time reversal symmetry, the growing surface at 
deposition probability $p$ should be identical to the inverted eroding surface at $1-p$.
In the monomer version ($Q=1$), it is well known that 
the unbalance ($p\neq q$) becomes a relevant perturbation to the EW fixed point and 
drives the system into the KPZ universality class. In this section, we 
investigate the scaling property of the nonequilibrium growing/eroding surfaces
for $Q\ge 2$. 

The nonequilibrium surfaces for the $Q$-mer model and the $Q$-particle-correlated model
show completely distinct  characteristics in both dynamic evolution and stationary morphology.
The former ones always facet, while the latter ones form global grooved structures.
For clarity, we discuss these two models separately in this section. 

\subsection{$Q$-mer model}

The nonequilibrium surfaces of the $Q$-mer model display sharply faceted structures.
Fig.~5 shows the time evolution of surfaces for the dimer model at $p=0.6$ and 0.1
in a typical simulation sample. In the growing surfaces, the faceting is 
caused by the spontaneous formation of pinning valleys. For the dimer model, 
any flat segment with odd size acts as the nucleus of a pinning valley. 
Such valleys can not be filled up by deposition of dimers and one empty blocked site 
always remain at the valley bottom. 
Pinning valleys can annihilate only in pairs, and only if their valley bottoms 
are at the same height. Therefore, the only way to remove pinning valleys 
and grow the surface is the evaporation of the entire hill between two pinning valleys.

In finite systems, the surface moves like shock waves. The initially flat surface evolves
into a faceted shape with only two remaining pinning valleys. 
The valleys are sharp and the hilltops are rounded  with a typical size $\xi_0$. 
Approaching the equilibrium point, $\xi_0$ diverges and the facets disappear.
The annihilation time of this last pair scales exponentially with system size. 
After its annihilation, the surface grows into another faceted shape very fast
and the whole evolution repeats itself. Details of this motion have been reported in 
Ref.~\cite{NPD}. 

For general $Q$-mer models, there exist $Q-1$ different kinds of pinning valleys 
with $n$  empty blocked sites at the valley bottom $(n=1,2, ..., Q-1)$. Two pinning 
valleys with $n$ and $m$ empty blocked sites merge into a 
pinning valley with $n+m$ empty blocked sites. When $n+m$ is a multiple of $Q$,
the valley can be filled up and the surface can grow. Even though the surface evolution
details are different for each $Q$, the basic evolution characteristics are equivalent,
i.e., a repeated jerky motion of locking into a facet shape and sudden evolving into another
facet shape. Time scale of this jerky motion is dominated by exponentially long
duration of a facet shape. If one starts with a flat surface of size $L$ that is not a multiple
of $Q$, at least one pinning valley always survives and the surface is pinned 
to the initial position.

The interface width $W$ of the faceted surface in Fig.~5 should be proportional to the
substrate size $L$. Therefore, the stationary roughness exponent is trivially $\alpha=1$. 
Effective exponents $\alpha_{eff}$ for the $Q$-mer model are plotted against 
$1/L$ in Fig.~6. As expected, $\alpha_{eff}$ converge to 1 for all $Q$. For small sizes,
finite size effects are present due to finite duration time of the facet shape. 

The early time growing behavior also can be predicted from annealing dynamics of
pinning valleys. Starting from a flat surface, the surface grows fast at the beginning
and many pinning valleys appear. The presence of pinning valleys slows down the growth
and the surface may grow only when two pinning valleys merge together by annealing out
the hill between them. As seen in Fig.~5, only a small region on the hilltops can be active
by deposition/evaporation processes. Near the valley, the ramps are inactive due to
the RSOS constraint. So, the annealing time of the hill should scale exponentially
with lateral size of the hill. After this time scale, the surface advances fast locally
up to the order of the hill height. In the faceted structure, the hill height should be 
proportional to the hill size. In this point of view, one can conclude that the local 
surface fluctuation should grow logarithmically with time. As any other time scale 
is present, we expect that $W\sim \log t$ in the early time regime~\cite{NPD,Gryn}.

In Fig.~7, we plot $W$ against $\log t$. For all $Q$, it shows a nice linear behavior.
For clarity, we assume that $W\sim [\log t]^\chi$ and plot effective exponents 
$\chi_{eff}$, defined similarly in Eq.~(\ref{beta}), versus $t$ in the inset.
They converge rather slowly but nicely to 1. We estimate that $\chi=1.1(1)$.

\subsection{$Q$-particle-correlated model}

Surface morphology for the $Q$-particle-correlated model 
is very surprising. Fig.~8 shows the time evolution of surfaces for
$Q=2$ at $p=0.6$ and 0.1 in a typical simulation sample. 
In contrast to the $Q$-mer model, the surfaces are not faceted but show
a global grooved structure with considerable local fluctuations. 

Starting from a flat surface, small scale hills and valleys start to form
and merge into a few macroscopic hills and valleys. These
macroscopic hills/valleys contain considerable roughness in microscopic
scale. Especially the hilltops of the growing surface are broad and rough 
with high activity of deposition/evaporation processes. In contrast, 
the valley bottoms are rather localized and their ramps are quite steep
without much activity.  However, they are not extremely localized or
sharp as in the faceted structure of the $Q$-mer model. 
The valley bottoms do not last long exponentially with system size, but only
algebraically. Therefore, there is no well-defined repeated motion of the
growing surface. A few macroscopic valleys merge and split stochastically
in the steady state regime. 

In Fig.~9, we plot $\alpha_{eff}$ against $1/L$
for $Q=2, 3,$ and 4. We estimate that $\alpha=1.00(5)$ for all $Q$. 
In contrast to the Q-mer model, the early time growth behavior for  
the $Q$-particle-correlated model is governed by a power law
with exponent $\beta$ varying with $Q$ and $p$.
We  plot $\beta_{eff} $ against $t$ at $p=0.6$  in Fig.~10. Our estimates
at $p=0.6$ are $\beta=0.460(1)$, 0.325(1), and 0.214(2) for $Q=2, 3$ and 4. 
The value of $\beta$ increases with increasing $p$ (or deposition/evaporation 
bias). The power-law type growth is consistent with our groove formation picture.
The ramps near the valley are still active, so the annealing time of 
the hill between two valleys does not scale exponentially but algebraically.

A similar grooved structure could be found in the so called conserved RSOS 
(CRSOS) model \cite{KIKJ} and other models with roughness exponent 
$\alpha\ge 1$ \cite{dyn}.  It may be quite interesting to investigate
similarities and differences among these models with grooved structures.
For example, the grooved structure in the CRSOS model displays a sharp
peak and a rounded valley, which is just the reverse of that  in our model.
Moreover, the physical origin for occurrence of macroscopic grooves
is not well explored. More detailed investigation is left for the future
study.

\section{Summary and Discussion}

We investigated the scaling properties of the interface fluctuation width for
the $Q$-mer and $Q$-particle-correlated deposition-evaporation models.
Both models are constrained with a global conservation law
that the particle number is conserved modulo $Q$ at each height level. 
A 1D surface of these models can be mapped on a random walk trajectory 
with $Q$-visiting global constraint. 

When deposition and evaporation are balanced, one can obtain
Gibbs-type equilibrium surfaces in the stationary state.
Then, the dispersion of the $Q$-visiting random walks (QVRW) 
can translate into the stationary roughness of equilibrium surfaces. 
However, the $Q$-mer dynamics is not ergodic 
and can not generate all possible QVRW configurations. This non-ergodicity 
may generate huge corrections to scaling and hinder the numerical analysis
for the stationary roughness. Especially for the dimer model, we found a significant
overshooting of roughness exponent $\alpha$ in finite-size-scaling analysis. 
In this paper, we introduce the $Q$-particle-correlated model that fully restores
the ergodicity. Our numerical study showed that the overshooting is fairly reduced
and our estimate for $\alpha$ is fully consistent with the QVRW result.

We also studied the time-dependent behavior in the early time regime. 
Interestingly, the growth exponent $\beta$ varies with models and $Q$
(see Table 1), in contrast to the robustness of $\alpha$. Therefore, 
these models correspond to a series of new universality classes. 

When the evolution dynamics is biased, the surface can grow or erode indefinitely
with time. As the detailed balance is violated, the steady-state surface configurations 
do not form a Gibbs-type ensemble. We numerically investigated the scaling properties of
these nonequilibrium surfaces. The $Q$-mer models display shock-wavelike evolutions
and finally form a macroscopic facet, while the $Q$-particle-correlated models
follow an ordinary power-law type growth and form a macroscopic grooved structure.
Like in the equilibrium cases, the steady state roughness do not depend on $Q$, but
the growth exponent for the $Q$-particle-correlated models varies with $Q$ and the
deposition/evaporation bias. These values are listed in Table 1. 

In summary, we found a series of new universality classes in deposition-evaporation models
with a global conservation law. Even though the stationary roughness in equilibrium
can be explained in terms of QVRW models, the growth diversity is not fully understood.
The scaling properties for growing/eroding surfaces for the $Q$-particle-correlated models
are surprising and wait for a reasonable intuitive explanation. Generalization to higher
dimensions are under current investigation.

\begin{acknowledgements}
We thank Jae Dong Noh for useful discussions.
This research is supported in part by grant No.~R01-2001-00025 (YK) and 
by grant No.~2000-2-11200-002-3 (HP) from the Basic Research Program of KOSEF.
\end{acknowledgements}

\end{multicols} 

\begin{table}[ht]
\caption{Roughness exponent $\alpha$ and growth exponent $\beta$
for the $Q$-mer and $Q$-particle-correlated ($Q$-PC) models. For nonequilibrium
$Q$-PC models, $\beta$ also varies with $p$.}
\label{table1}
\begin{tabular}{cccccccc}
\multicolumn{1}{c}{}&
\multicolumn{1}{c}{Model}&
\multicolumn{3}{c}{$\alpha$}&
\multicolumn{3}{c}{$\beta$}\\ 
& & $Q=2$ & $Q=3$ & $Q=4$ & $Q=2$ & $Q=3$ & $Q=4$ \\ \hline
Equilibrium & $Q$-mer & 0.29(4) & 0.32(4) & 0.33(3) & 0.108(1) & 0.100(1) & 0.098(3)\\
$(p=q)$ & $Q$-PC & 0.32(3) & 0.33(2) & 0.32(4) & 0.194(1) & 0.140(2) & 0.097(2)\\ \hline
\multicolumn{1}{c}{Nonequilibrium} & 
\multicolumn{1}{c}{$Q$-mer} &
\multicolumn{3} {c}{1.00(5) \quad\quad (facet)} &
\multicolumn{3} {c}{ 0 \quad $(\log t)$ } \\
\multicolumn{1}{c}{$(p=0.6)$} & 
\multicolumn{1}{c}{$Q$-PC} &
\multicolumn{3} {c}{1.00(5) \quad (groove)} &
\multicolumn{1} {c}{ 0.460(1) } &
\multicolumn{1} {c}{ 0.325(1) } &
\multicolumn{1} {c}{ 0.214(2) } \\ 
\end{tabular}
\end{table}

\newpage
\begin{figure}
\centerline{\epsfxsize=130 mm  \epsfbox{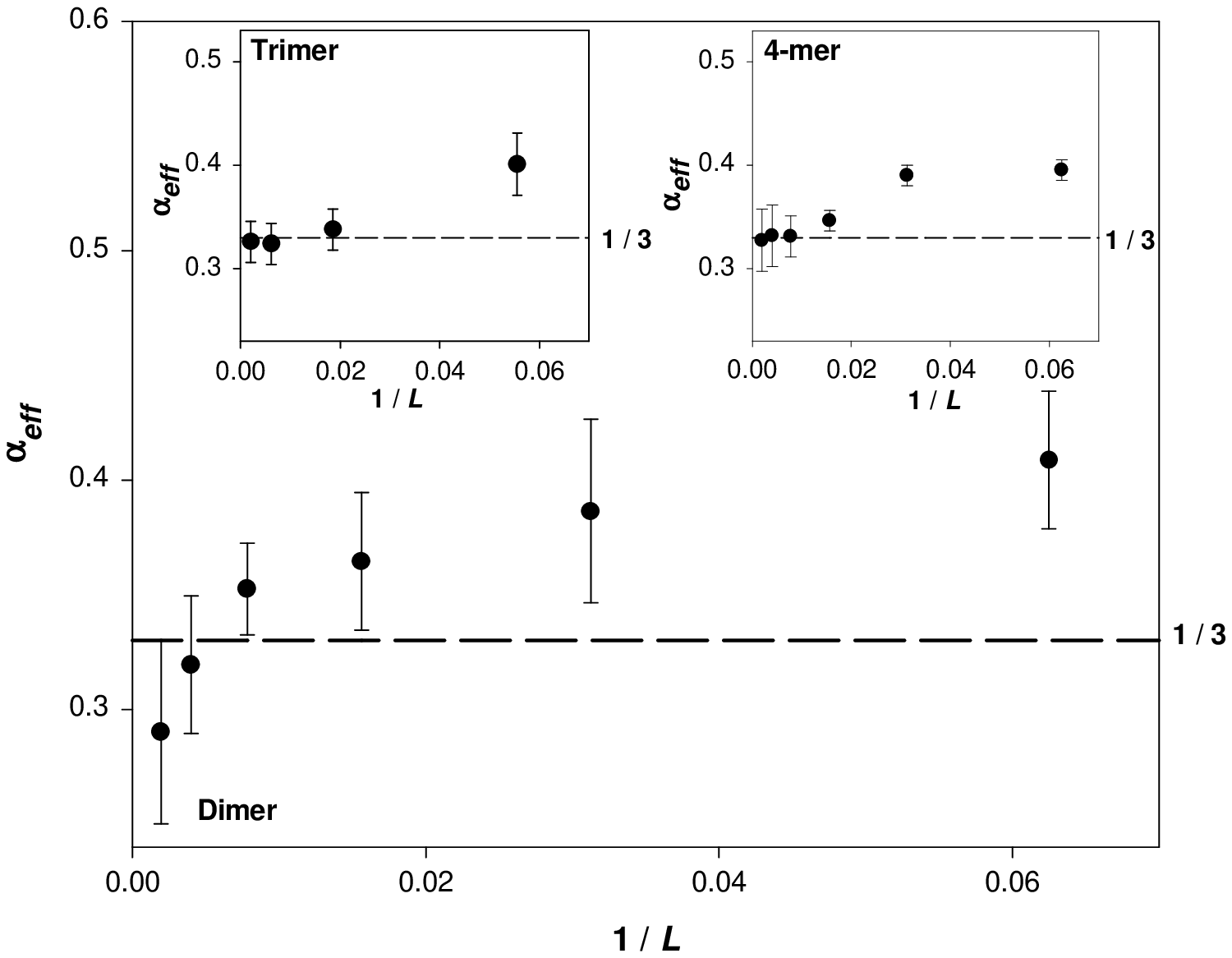}}
\vskip 10 true pt
\caption{Effective stationary roughness exponents $\alpha_{eff}$ versus $1/L$ for the $Q$-mer 
deposition-evaporation models at $p=1/2$(equilibrium surfaces).
The horizontal broken lines represent $\alpha_{eff}=1/3$. }
\end{figure} 
 
\begin{figure}
\centerline{\epsfxsize=130 mm  \epsfbox{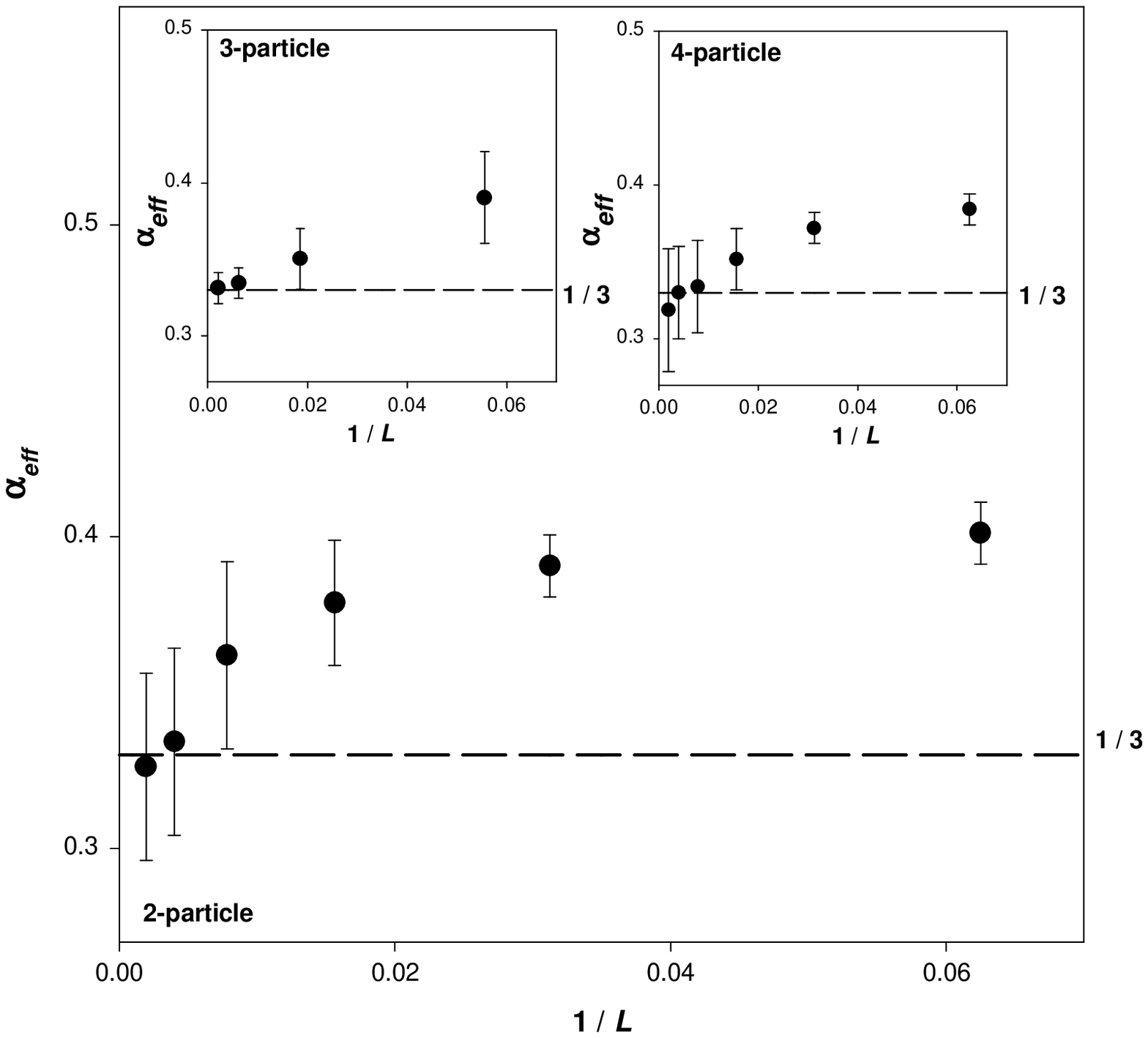}}
\vskip 10 true pt
\caption{ Effective stationary roughness exponents $\alpha_{eff}$
versus $1/L$ for the $Q$-particle-correlated deposition-evaporation models at $p=1/2$.
The horizontal broken lines represent $\alpha_{eff}=1/3$. }
\end{figure}

\begin{figure}
\centerline{\epsfxsize=130 mm  \epsfbox{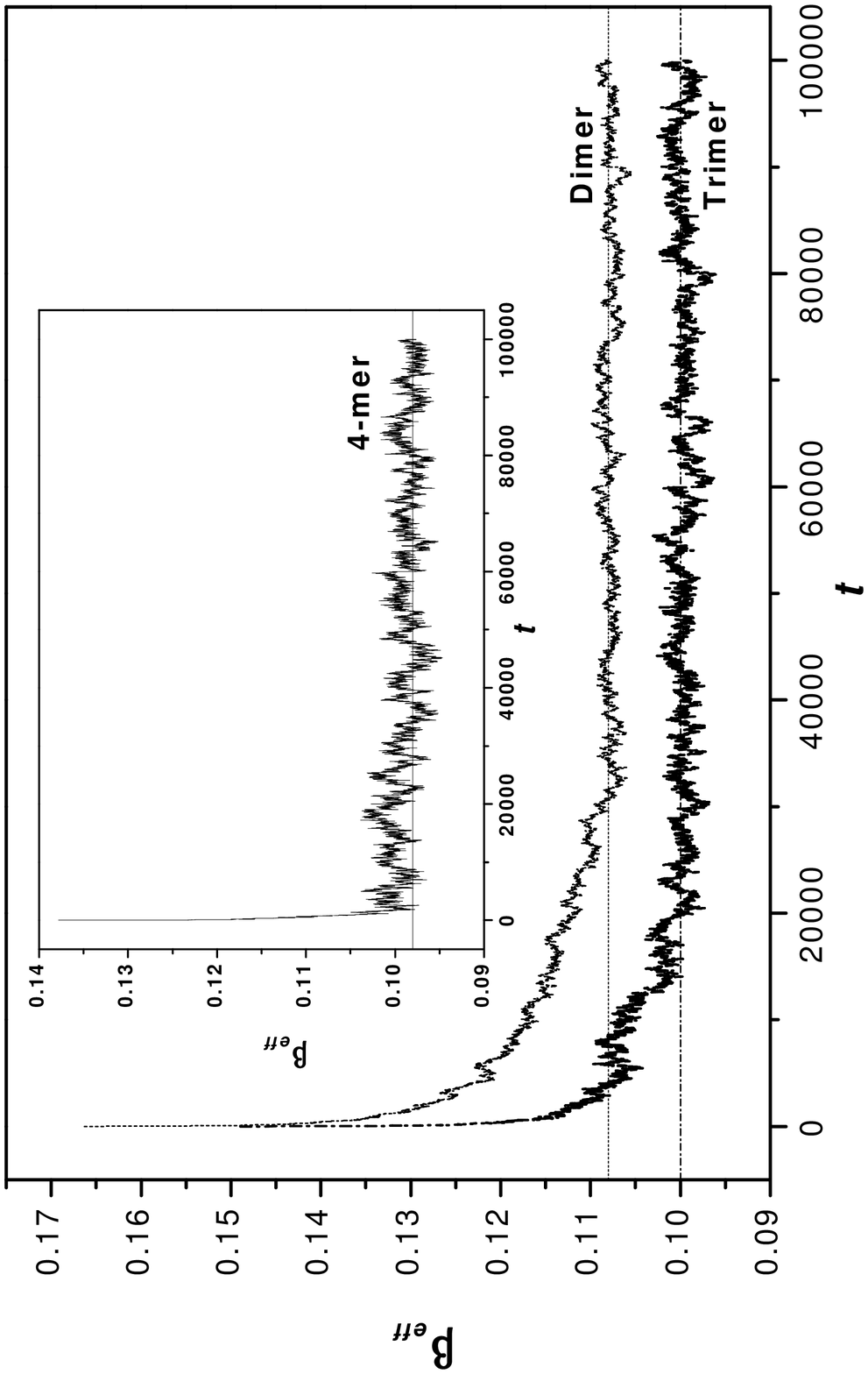}}
\vskip 10 true pt
\caption{Effective growing exponents $\beta_{eff}$ versus time $t$ 
for the $Q$-mer models at $p=1/2$.  
The horizontal lines are $\beta_{eff}=0.108$, 0.100, and 0.098 for 
$Q=2, 3$, and 4,  respectively. }
\end{figure}

\begin{figure}
\centerline{\epsfxsize=130 mm  \epsfbox{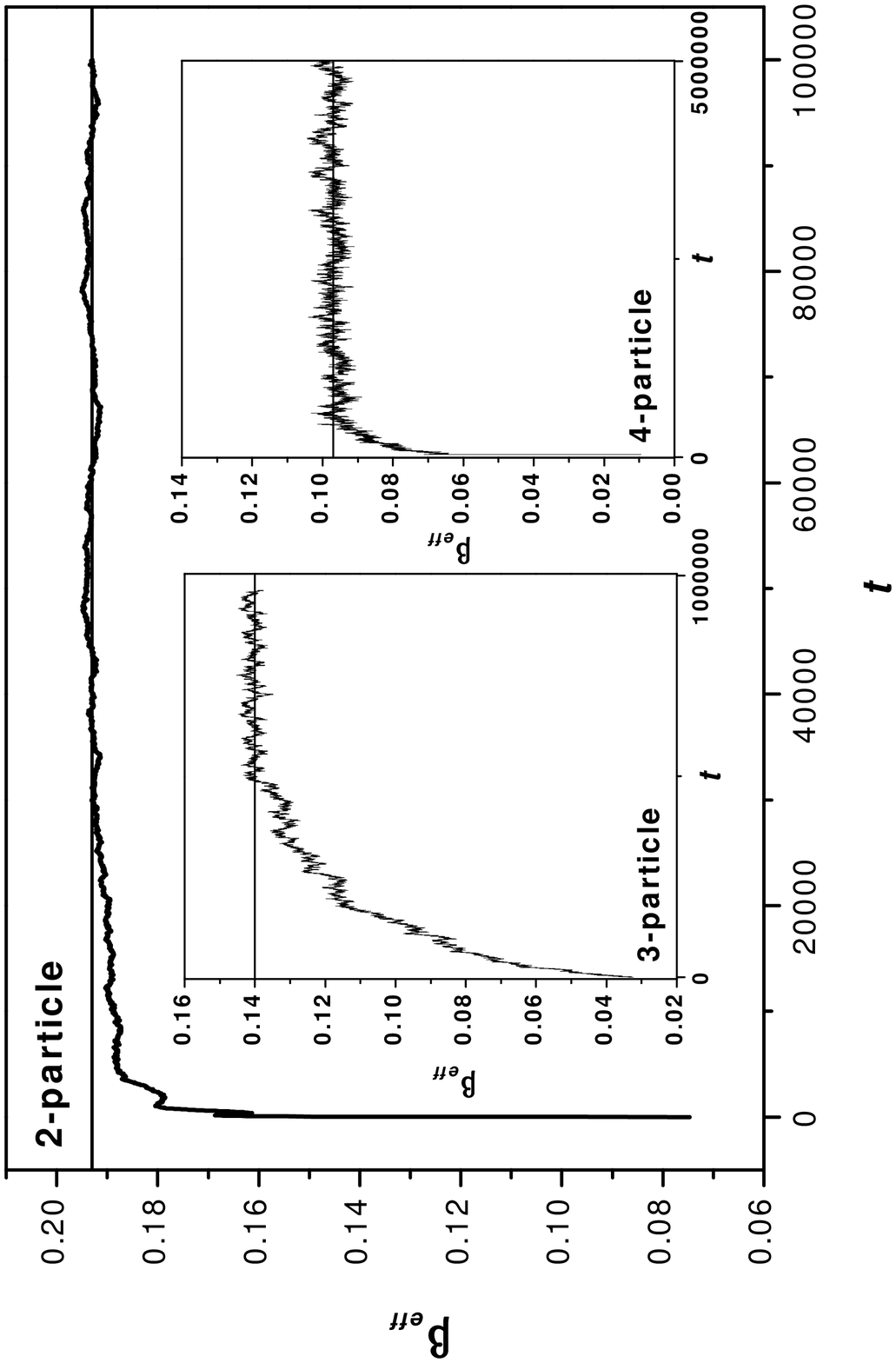}}
\vskip 10 true pt
\caption{Effective growing exponents $\beta_{eff}$ versus time $t$ 
for the $Q$-particle-correlated models at $p=1/2$.
The horizontal lines are $\beta_{eff}=0.194$, 0.140, and 0.097 for $Q=2$, 3, and 
4, respectively. 
We take a rather small substrate size $L=10^3$ for $Q=4$,
due to huge initial transient behavior. }
\end{figure}

\begin{figure} 
\centerline{\epsfxsize=130 mm  \epsfbox{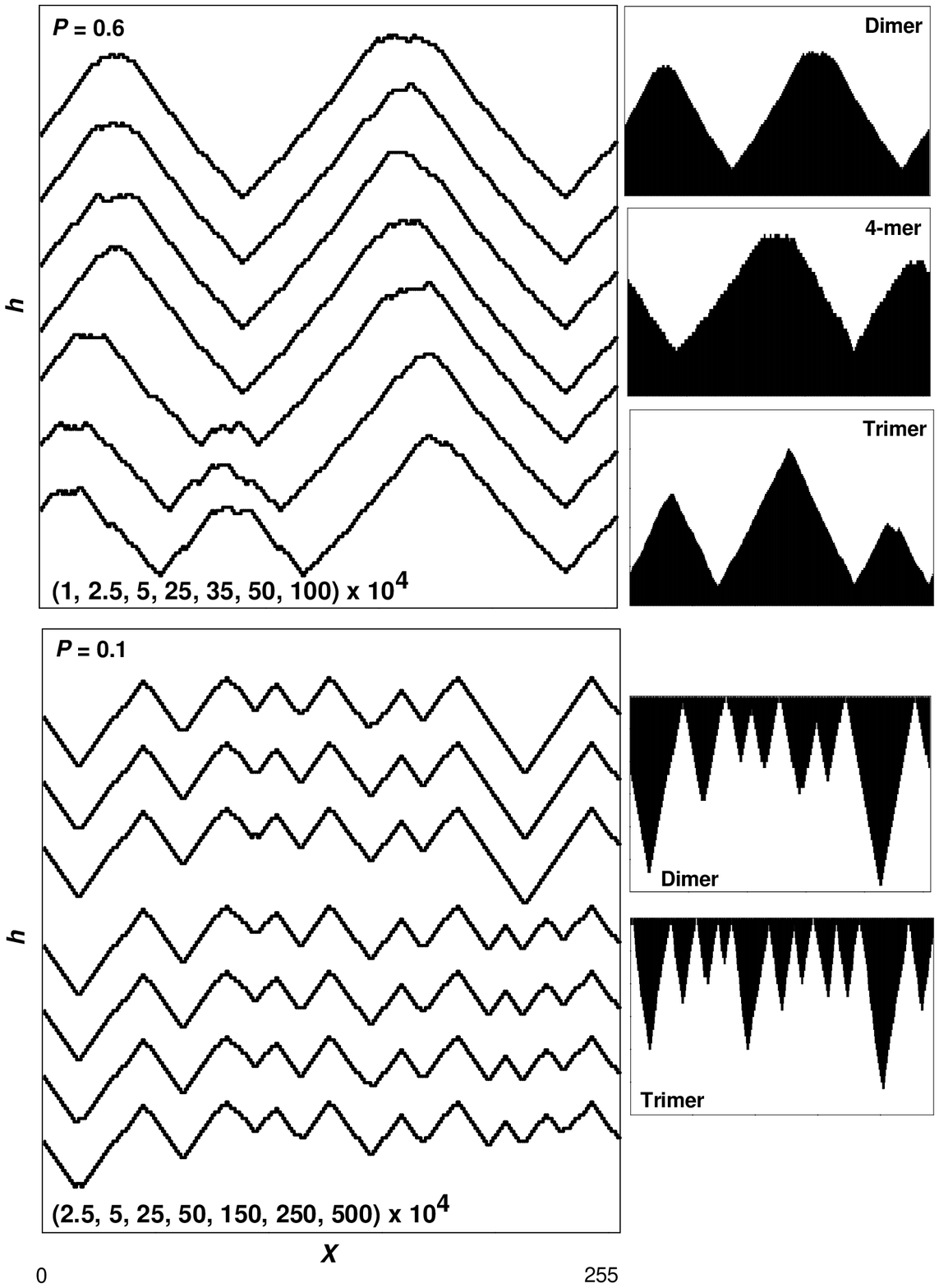}}
\vskip 10 true pt
\caption{Time evolutions of surfaces
for dimer growth model at $p=0.6$ and $p=0.1$
in a typical simulation sample. 
The numbers in the bottom of 
figures denote Monte Carlo times when each surface configuration
is taken. Typical surface configuration with maximum interface width
are drawn in the right for $Q=2, 3$, and 4.}
\end{figure} 

\begin{figure}
\centerline{\epsfxsize=130 mm  \epsfbox{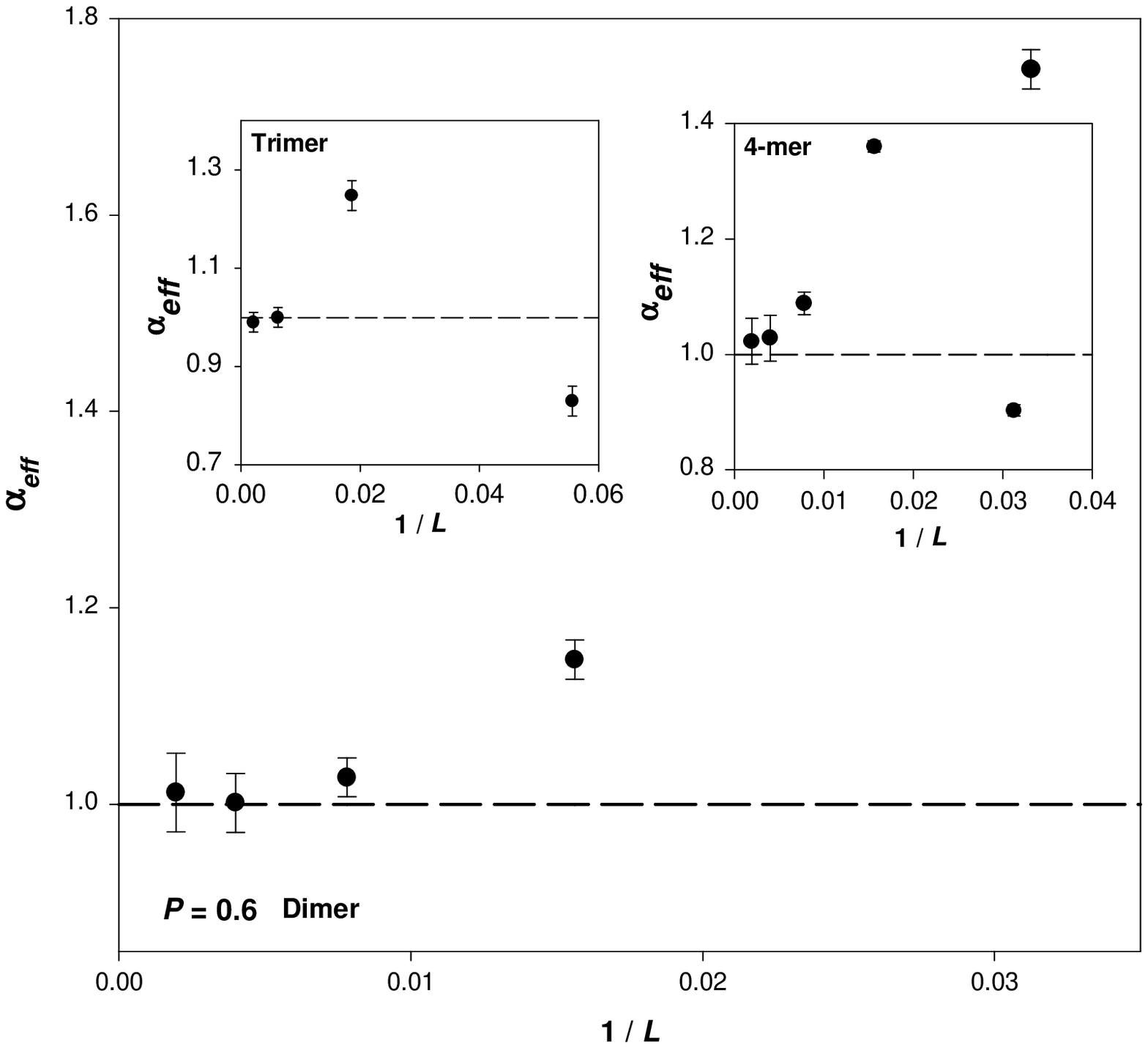}}
\vskip 10 true pt
\caption{ Effective stationary roughness exponents $\alpha_{eff}$
versus $1/L$ for the $Q$-mer models at $p=0.6$.
The horizontal broken lines represent $\alpha_{eff}=1$. }
\end{figure} 

\begin{figure}
\centerline{\epsfxsize=130 mm  \epsfbox{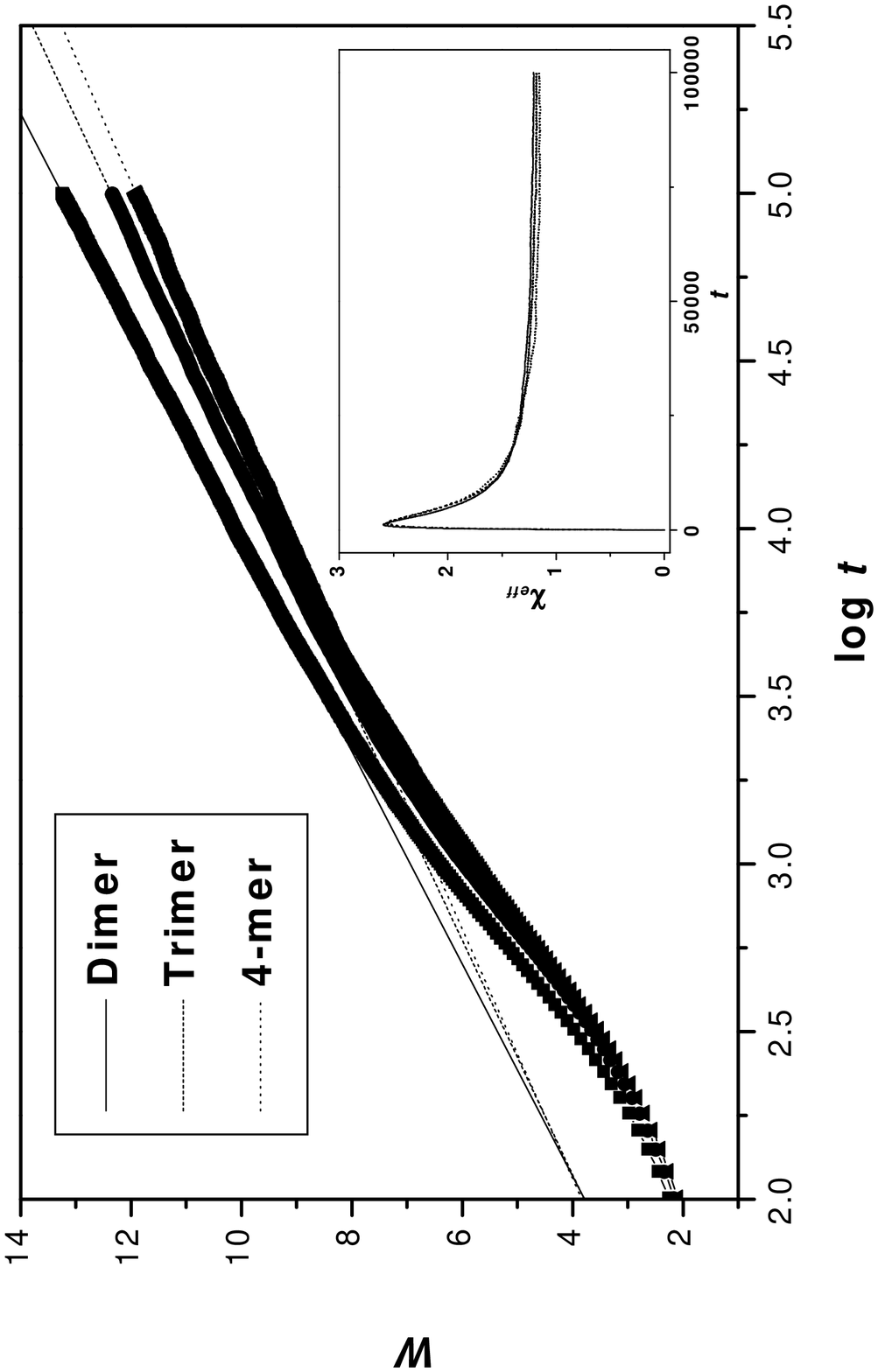}}
\vskip 10 true pt
\caption{Interface width $W$ against $\log t$ for the $Q$-mer models 
at $p=0.6$ in $t \ll L^{z_W}$. Assuming that $W\sim [\log t]^\chi$, 
we plot effective exponents $\chi_{eff}$ versus $t$ in the inset.
$\chi_{eff}$ converge rather slowly but nicely to 1. }
\end{figure}

\begin{figure} 
\centerline{\epsfxsize=130 mm  \epsfbox{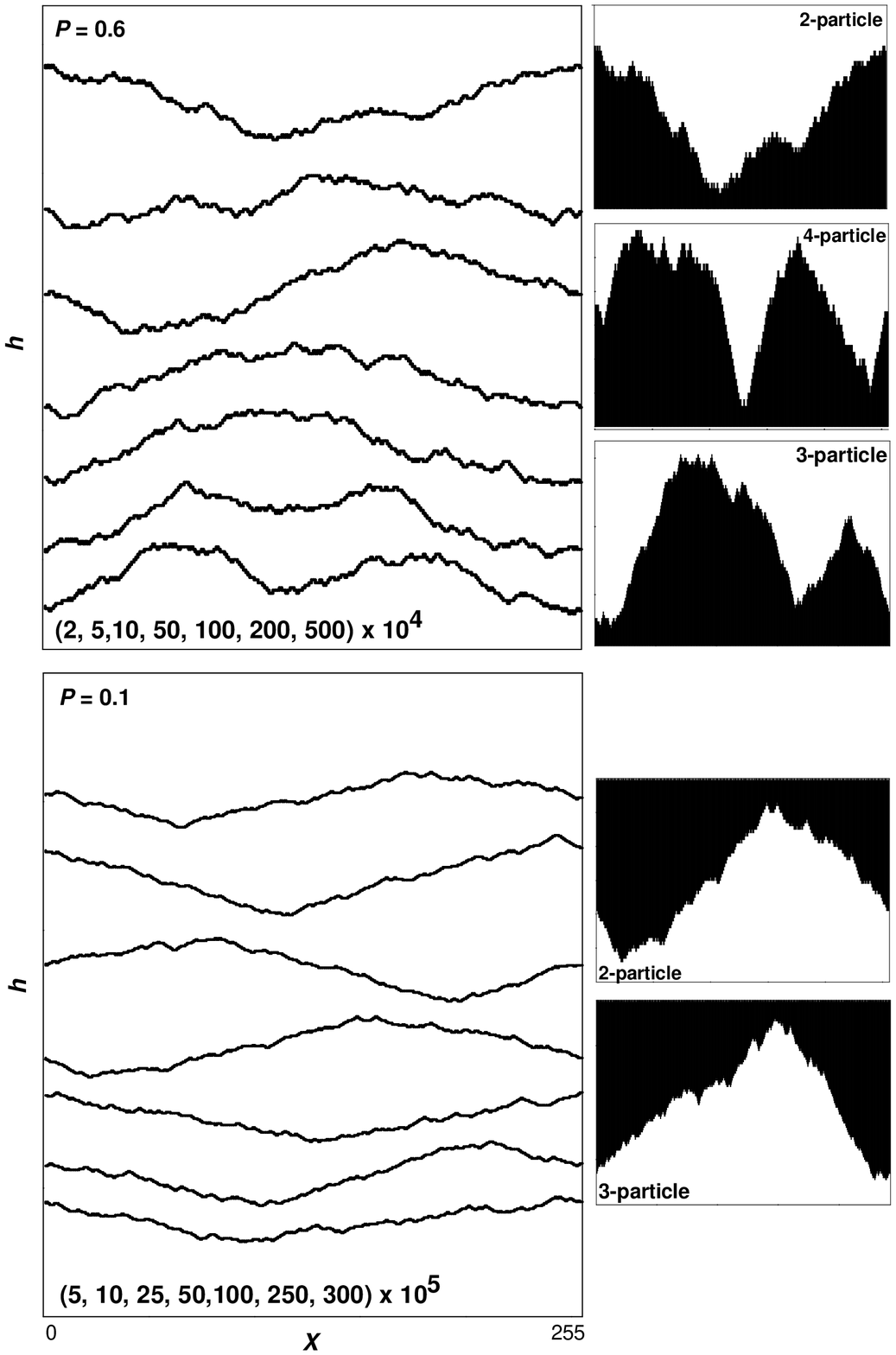}}
\vskip 10 true pt
\caption{Time evolutions of surfaces
for the $2$-particle-correlated model at $p=0.6$ and $p=0.1$
in a typical simulation sample. 
Typical surface configuration with maximum interface width
are drawn in the right for $Q=2, 3$, and 4.}
\end{figure} 

\begin{figure}
\centerline{\epsfxsize=130 mm  \epsfbox{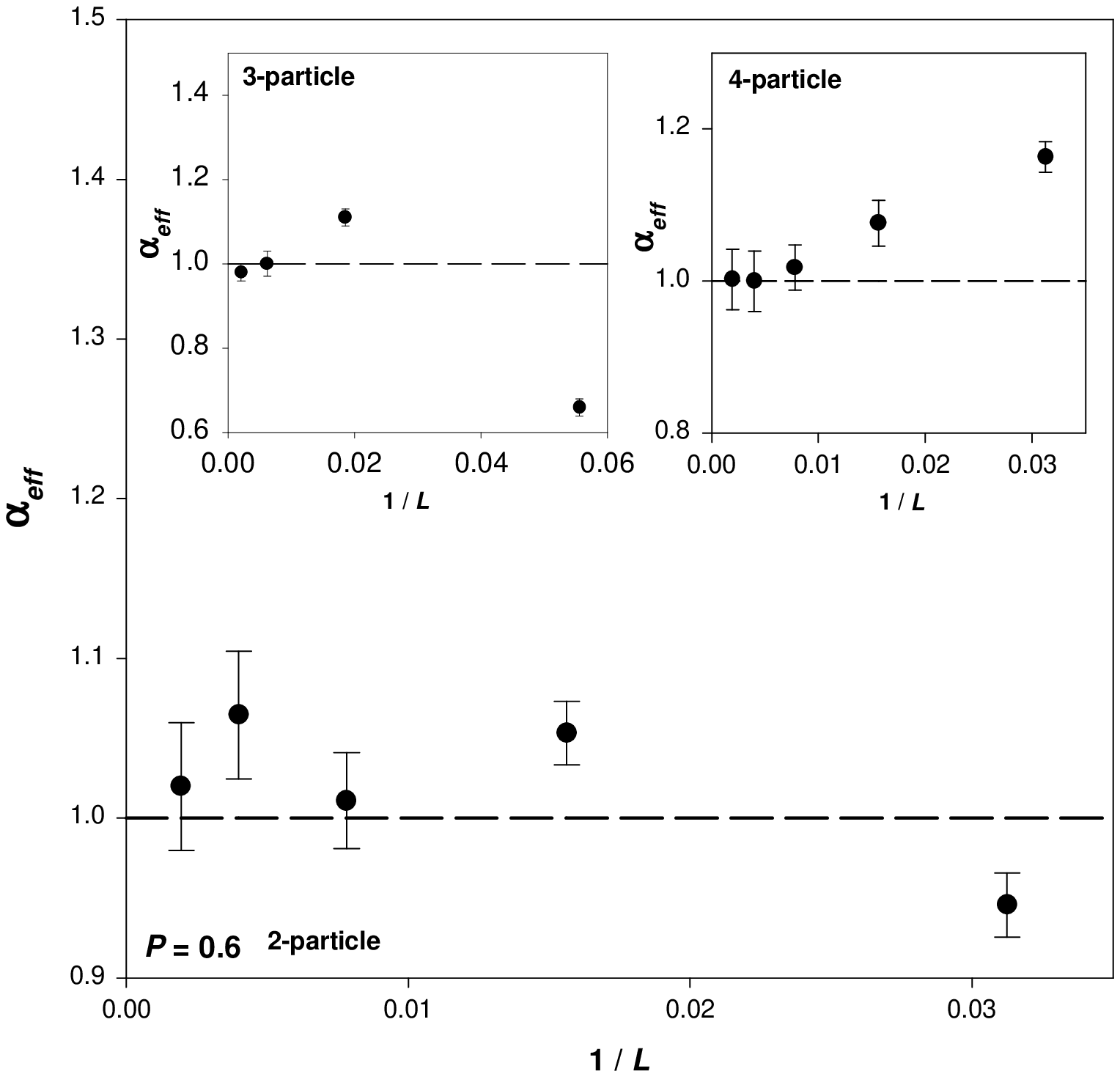}}
\vskip 10 true pt
\caption{ Effective stationary roughness exponents $\alpha_{eff}$
versus $1/L$ for the $Q$-particle-correlated models at $p=0.6$.
The horizontal broken lines represent $\alpha_{eff}=1$. }
\end{figure} 
 
\begin{figure}
\centerline{\epsfxsize=130 mm  \epsfbox{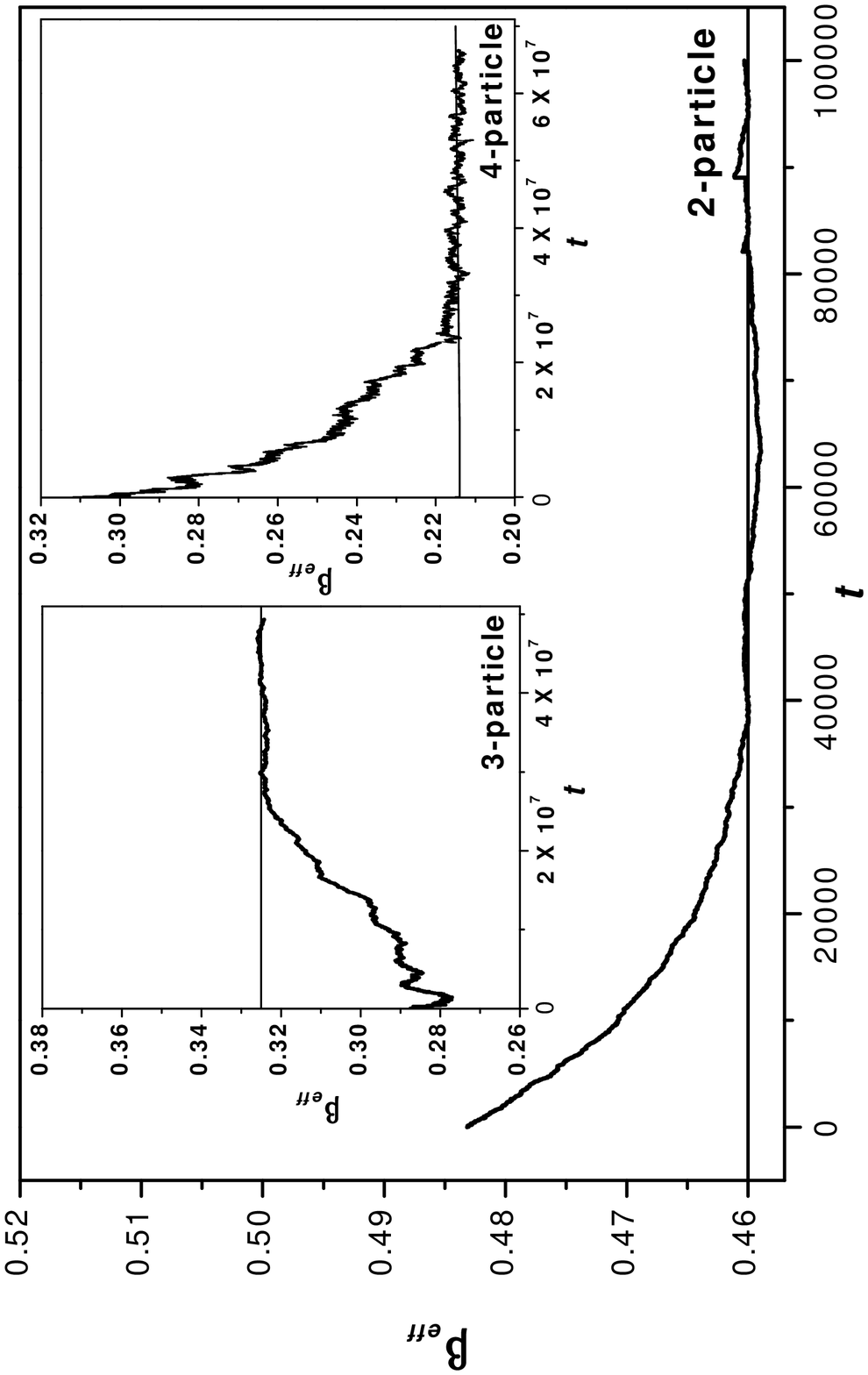}}
\vskip 10 true pt
\caption{Effective growing exponents $\beta_{eff}$ versus time $t$ 
for the $Q$-particle-correlated models at $p=0.6$.
The horizontal lines are $\beta_{eff}=0.460$, 0.325, and 0.214 for $Q=2$, 3, and 
4, respectively. } 
\end{figure}

\end{document}